\documentclass[prb,amsmath,amssymb,twocolumn,showpacs,floatfix]{revtex4}
\usepackage{graphicx}
\usepackage[english]{babel}
\usepackage{times}
\usepackage{dcolumn}
\usepackage[usenames,dvipsnames]{color}
\usepackage{units}

\begin{document}

\title{Quantum impurity on the surface of a topological insulator}

\author{Rok \v{Z}itko}

\affiliation{Jo\v{z}ef Stefan Institute, Jamova 39, SI-1000 Ljubljana,
Slovenia}

\date{\today}

\begin{abstract}
  It is shown that the Hamiltonian for a quantum magnetic impurity on
  the surface of a topological insulator can be mapped to the
  conventional pseudo-gap Anderson impurity model, albeit the
  combinations of the continuum states which hybridize with the impurity
  have more complex structure in the reciprocal and spin space. If
  the Fermi level is away from the Dirac point, the impurity is
  predicted to be fully screened at low enough temperatures, i.e.,
  there are no residual degrees of freedom.
\end{abstract}


\maketitle

\newcommand{\vc}[1]{{\mathbf{#1}}}
\newcommand{\vck}{\vc{k}}
\newcommand{\braket}[2]{\langle#1|#2\rangle}
\newcommand{\expv}[1]{\langle #1 \rangle}
\newcommand{\ket}[1]{| #1 \rangle}

Due to the spin-orbit coupling, insulators in the class of
``topological insulators'' have non-trivial topological properties of
the bulk band structure, which results in the presence of peculiar
metallic states on their surfaces \cite{fu2007, hsieh2009, moore2010}.
Recently studied materials such as Bi$_2$Se$_3$ and Bi$_2$Te$_3$ show
topological-insulator behavior even at high temperatures and have the
simplest allowed surface states: a single nondegenerate Dirac cone
\cite{xia2009, zhangH2009, chen2009}. These surface states have a
characteristic relation between the momentum and spin, which makes the
surface electrons insensitive to scattering by impurities since a
back-scattering event would require the electron spin to be flipped,
yet this is prohibited by the time-reversal symmetry
\cite{roushan2009, fu2009, alpichshev2010, zhangT2009, guo2010}.
Recent Fourier-transform scanning tunneling spectroscopy (FT-STS)
experiments have confirmed this prediction \cite{roushan2009,
alpichshev2010, zhangT2009}. Such topological insulator materials may
find applications in spintronics \cite{chen2009} and quantum computing
\cite{fu2008}.

The surface-state electrons on topological insulators have definite
chirality and for the non-degenerate states near the Dirac point one
has $\expv{ \vc{s}(-\vck) }=-\expv{ \vc{s}(\vck) }$ \cite{fu2007,
zhangT2009, guo2010}. Such surface states are expected to be perturbed
differently by non-magnetic and magnetic impurities. The
time-reversal-symmetry breaking perturbations allow scattering
between the time-reversed states $\ket{\vck,\uparrow}$ and
$\ket{-\vck,\downarrow}$ \cite{guo2010}. The case of classical spins
(which clearly break the time-reversal symmetry) has been explored in
Ref.~\onlinecite{liu2009}, where it was predicted that the impurity
opens up a local gap and suppresses the local density of states.
Scattering on static spin-dependent impurity potential was also
addressed in Refs.~\onlinecite{guo2010} and \onlinecite{zhou2009},
where characteristic interference features were predicted to be
observable by spin-polarized FT-STS. The description of a magnetic
impurity in terms of classical/static spin is, however, a rather crude
approximation; magnetic impurities are dynamic objects and should be
described using quantum impurity models \cite{bulla2008}. For
impurities described, for example, by the Anderson impurity model
\cite{anderson1961}, the local moment of a quantum impurity may be
compensated at low temperatures by the Kondo effect \cite{kondo1967,
nozieres1974, wilson1975, andrei1983, bulla2008}. Kondo screened
impurities are effectively non-magnetic, thus the extent of
backscattering on magnetic adsorbates will critically depend on the
degree of screening of the local moment by the surface-state electrons
\cite{zhangT2009}. Due to the spin-orbit coupling, there is no SU(2)
symmetry in the spin sector alone \cite{fu2009}, therefore it is not
immediately clear how effectively the impurity spin can be screened by
the chiral (or helical) surface-state electrons. In
Ref.~\onlinecite{feng2009}, this problem has been approached using the
Varma-Yafet variational Ansatz. Only partial screening (one third)
of the local moment was found and $1/T$ Curie-type magnetic
susceptibility behavior was predicted at low temperatures. 
In variational methods, there are no a-priori guarantees that the
correlation functions are correct. In addition, the variational Ansatz
which projects out the double occupancy of the impurity might itself
be problematic. For this reason, the deduction of the one-third
screening from the spin-spin correlation functions is uncertain. In
this work, this problem is reconsidered by a different approach,
showing that the effective quantum impurity model actually takes a
very simple form of the conventional single-channel Anderson impurity
model, even though the nature of the hybridizing combinations of the
conduction-band states is indeed non-trivial. For the generic
situation with the Dirac point away from the Fermi level, the local
moment will be screened below the Kondo temperature $T_K$, and the
impurity would become effectively non-magnetic. No anomalies are
expected in the low-temperature thermodynamics in this case. Since the
experimental studies of magnetically doped topological insulators are
already under way \cite{cha2010}, it is important to arrive at better
understanding of the Kondo effect in the case of chiral surface state
electrons.

In the simplest case, the surface-state electrons on the surface of a
topological insulator may be described by the Hamiltonian \cite{liu2009}
\begin{equation}
\label{ham}
\begin{split}
H_0 &= \sum_{\vck,\alpha,\beta} \psi^\dag_{\vck\alpha}
h_{\alpha\beta}(\vck)\psi_{\vck\beta}, \\
h_{\alpha\beta}(\vck) &= \hbar v_F \left(
k_x \sigma^x_{\alpha\beta}
+k_y \sigma^y_{\alpha\beta}
\right),
\end{split}
\end{equation}
where $\vck=\{k_x,k_y\}$ is a two-dimensional (2D) vector in the
reciprocal space of the surface-state band, $\alpha$ and $\beta$ are
spin indexes ($\uparrow$ or $\downarrow$), $v_F$ is the Fermi
velocity, and $\sigma^{x,y}$ are the Pauli matrices.  \footnote{The
general conclusions of this work are also valid for a ``helical''
surface-state Hamiltonian with $h(\vck) = \hbar v_F (k_x \sigma^y-k_y
\sigma^x)$, where the spins are rotated by 90 degrees around the
$z$-axis as compared to Eq.~\eqref{ham}, as already noted in
Ref.~\onlinecite{feng2009}.} The $\sigma$ matrix is proportional to
real spin \cite{liu2010}. We write $\vck=k \{ \cos\phi, \sin\phi \}$
where $k=|\vc{k}|$, and we introduce the energy $\epsilon_k = \hbar
v_F k$. We write
\begin{equation}
\label{eqM}
M_{\alpha\beta}(\phi) \equiv \left( \cos\phi\ \sigma^x + 
\sin\phi\ \sigma^y \right)_{\alpha\beta} = \begin{pmatrix} 0 & e^{-i\phi} \\
e^{i\phi} & 0 \end{pmatrix}
\end{equation}
and thus $h_{\alpha\beta}(\vck) = \epsilon_k M_{\alpha\beta}(\phi)$.
The hybridization term is 
\begin{equation}
H_\mathrm{hyb} = \sum_{\vck,\alpha} V_{\vck\alpha}
d^\dag_\alpha \psi_{\vck\alpha} +
V_{\vck\alpha}^*
\psi^\dag_{\vck\alpha} d_\alpha,
\end{equation}
where $d^\dag$ and $d$ are the creation and annihilation operator for
the impurity electron. In this work we take $V_{\vck\alpha}=V_\vck$;
this is an approximation, since the surface states labeled by the
spin index $\sigma$ are not eigenstates of $S_z$. The impurity
Hamiltonian is simply
\begin{equation}
H_\mathrm{imp} = \sum_\alpha \epsilon\ n_\alpha
+ U n_\uparrow n_\downarrow.%
\end{equation}
where $n_\alpha=d^\dag_\alpha d_\alpha$ is the occupancy of the spin
$\alpha$ level,
$\epsilon$ is the impurity energy level, and $U$ the on-site
electron-electron repulsion.

The reduction of the problem to an effective one-dimensional quantum
impurity problem is analogous to the derivations in
Ref.~\onlinecite{krishna1980a}, but performed for the 2D case. We go
from the discrete $\vc{k}$ to the continuous vectors $\vc{k}$ in the
standard way, $\frac{1}{N} \sum_\vc{k} \to \frac{1}{(2\pi)^2} \int
d^2\vck$,
where $N$ is the number of the surface states, and the components of
the wave vector on the right-hand side run from $-\pi$ to $\pi$. The
continuum operators that correspond to $\psi_{\vck\alpha}$ are
$a_{\vck\alpha}$ and they are normalized as 
$\{ a_{\vck\alpha}, a^\dag_{\vck'\alpha'} \}=\delta(\vck-\vck')
\delta_{\alpha,\alpha'}$,
thus the mapping of the operators is $\psi_{\vck\alpha} \to
(2\pi/\sqrt{N}) a_{\vck\alpha}$. This leads to
\begin{equation}
\begin{split}
H_0 &= \sum_{\alpha,\beta} \int d^2\vck\ a^\dag_{\vck\alpha} h_{\alpha\beta}(\vck) 
a_{\vck\beta}, \\
H_\mathrm{hyb} &= \sum_{\alpha} \frac{\sqrt{N}}{2\pi} 
\int d^2\vck\ V_\vck d^\dag_\alpha a_{\vck\alpha} + \mathrm{H.c.}
\end{split}
\end{equation}

In the next step we expand the operators $a_{\vck\alpha}$ in the
azimuthal components. In fact, we allow for a slightly more general Ansatz:
\begin{equation}
\label{eq9}
a_{\vck\alpha} = \frac{1}{\sqrt{k}} \frac{1}{\sqrt{2\pi}} \sum_{m,\beta} e^{im\phi} U_{\alpha\beta}(\phi)
c_{km\beta},
\end{equation}
where $U$ is some unitary matrix which may depend on the azimuthal
angle $\phi$, while $m \in \mathbb{Z}$. The operators $c_{km\beta}$
are normalized as 
\begin{equation}
\{ c_{km\beta}, c^\dag_{k'm'\beta'} \} = \delta(k-k') \delta_{m,m'}
\delta_{\beta,\beta'}.
\end{equation}
The inverse transformation is 
\begin{equation}
c_{km\beta} = \sqrt{k} \frac{1}{\sqrt{2\pi}} \int_0^{2\pi} d\phi\, e^{-im\phi}
\sum_\alpha U^\dag_{\alpha\beta}(\phi) a_{\vck\alpha}.
\end{equation}

The matrix $M$ in Eq.~\eqref{eqM} may be diagonalized using the
unitary matrix
\begin{equation}
W = \begin{pmatrix} 0 & e^{-i\phi/2} \\ e^{i\phi/2} & 0 \end{pmatrix}
\cdot e^{-i \sigma^y \pi/4}
= \frac{1}{\sqrt{2}} \begin{pmatrix} e^{-i\phi/2} & e^{-i\phi/2}
\\ e^{i\phi/2} & -e^{i\phi/2} \end{pmatrix},
\end{equation}
which gives $W^\dag M W = \begin{pmatrix} 1 & 0 \\ 0 & -1
\end{pmatrix} = \sigma^z$.
The transformation matrix $W$ has period $4\pi$, which reflects that a
quantized Berry's phase of $\pi$ is acquired by an electron circling
the Fermi arc, as characteristic for a ``topological metal'' surface
phase where the Fermi arc encloses a single Dirac point \cite{fu2007,
xia2009}.

Choosing $U=W$ in Eq.~\eqref{eq9}, we obtain a simple diagonal form for
the band Hamiltonian:
\begin{equation}
H_0 = \sum_{m,\sigma} \int dk \epsilon_{k\sigma} 
c^\dag_{km\sigma} c_{km\sigma},
\end{equation}
with $\epsilon_{k\uparrow}=\epsilon_k$ and
$\epsilon_{k\downarrow}=-\epsilon_k$. 
The density of states is
\begin{equation}
\rho_\sigma(\epsilon) 
= \frac{N}{(2\pi)^2} \int d^2\vc{k} \delta(\epsilon-\epsilon_{k\sigma})
= \frac{N}{2\pi}\frac{1}{(\hbar v_F)^2} |\epsilon| \theta_\sigma(\epsilon),
\end{equation}
where $\theta_\uparrow(x)=\theta(x)$ and $\theta_\downarrow(x) =
\theta(-x)$, with $\theta(x)$ the Heaviside step function. Thus
$\sigma=\uparrow$ corresponds to the upper Dirac cone, and
$\sigma=\downarrow$ to the lower Dirac cone. 

For simplicity, we will at first assume the hopping constants $V_\vck$
to be isotropic in the 2D space, i.e., $V_\vck = V_k$. Such
$\vc{k}$-dependence corresponds, for example, to the $d_{3z^2-r^2}$
impurity orbital. The hybridization term then transforms as
\begin{equation}
\begin{split}
H_\mathrm{hyb} &= \frac{\sqrt{N}}{2\pi} \sum_{\alpha,\beta}
\int \sqrt{k} dk V_k \\
&\times \left[ \frac{1}{\sqrt{2\pi}} \int d\phi \sum_{m,\beta}
e^{im\phi} W_{\alpha\beta}(\phi) \right] 
d^\dag_\alpha c_{km\beta}
+\mathrm{H.c.}
\end{split}
\end{equation}
Noting that for integer $m$
\begin{equation}
\label{cpm}
\frac{1}{\sqrt{2}}\frac{1}{2\pi}
\int_0^{2\pi}d\phi\ e^{i(m\pm 1/2)\phi} = 
\frac{1}{\sqrt{2}}\frac{1}{2\pi} \frac{4i}{2m \pm 1}
\equiv \gamma_m^\pm,
\end{equation}
we obtain
\begin{equation}
\begin{split}
H_\mathrm{hyb} &= 
\frac{\sqrt{N}}{\sqrt{2\pi}} 
\sum_{m,\alpha,\beta}
\int \sqrt{k}\ dk\ V_k \\
&\times
\begin{pmatrix}
\gamma_m^- & \gamma_m^- \\
\gamma_m^+ & -\gamma_m^+
\end{pmatrix}_{\alpha\beta}
d^\dag_\alpha c_{km\beta} + \mathrm{H.c.}
\end{split}
\end{equation}

Finally, we introduce the energy representation by defining
$c_{\epsilon m\sigma}=(d\epsilon_{k\sigma}/dk)^{-1/2} c_{km\sigma}$,
with $\epsilon=\epsilon_{k\sigma}$. The normalization then becomes
$\{ c_{\epsilon m \sigma}, c^\dag_{\epsilon'm\sigma'} \}
= \delta(\epsilon-\epsilon') \delta_{m,m'} \delta_{\sigma,\sigma'}$.
Note that $\epsilon \geq 0$ for $\sigma=\uparrow$ and $\epsilon \leq
0$ for $\sigma=\downarrow$. We introduce an upper energy cutoff for
$\sigma=\uparrow$ at $+D$ (the lower limit is 0) and a lower energy
cutoff for $\sigma=\downarrow$ at $-D$ (the upper limit is 0). We then
obtain
\begin{equation}
H_0 = 
 \sum_m \int_0^D d\epsilon\ \epsilon\ c_{\epsilon m \uparrow}^\dag
c_{\epsilon m \uparrow} \\
+
 \sum_m \int_{-D}^0 d\epsilon\ \epsilon\ c_{\epsilon m \downarrow}^\dag
c_{\epsilon m \downarrow}.
\end{equation}

For the hybridization term we have
\begin{equation}
\begin{split}
\frac{\sqrt{N}}{\sqrt{2\pi}}
& \int \sqrt{k}\ dk\ V_k c_{km\beta} \\
&= \frac{\sqrt{N}}{\sqrt{2\pi}}
\int \sqrt{k}\ dk\ \left( \frac{d\epsilon_{k\beta}}{dk} \right)^{1/2}
V_k c_{\epsilon m\beta} \\
&= \frac{\sqrt{N}}{\sqrt{2\pi}}
\int \sqrt{k_\epsilon} d\epsilon \left( \frac{dk}{d\epsilon_{k\beta}} \right)^{1/2}
V(\epsilon) c_{\epsilon m\beta}\\
&=
\int d\epsilon [\rho_\beta(\epsilon)]^{1/2} V(\epsilon) c_{\epsilon
m\beta},
\end{split}
\end{equation}
since the density of states is $\rho_\sigma(\epsilon)=(N/2\pi)
k_\epsilon |dk/d\epsilon|$.
We conclude that
\begin{equation}
H_\mathrm{hyb} = \sum_{m,\alpha,\beta} \int d\epsilon\,
\sqrt{\rho_\beta(\epsilon)} V(\epsilon) 
\begin{pmatrix}
\gamma_m^- & \gamma_m^- \\
\gamma_m^+ & -\gamma_m^+
\end{pmatrix}_{\alpha\beta}
d^\dag_\alpha c_{\epsilon m\beta} + \mathrm{H.c.}
\end{equation}
The hybridization function is defined as $\Gamma_\sigma(\epsilon)=\pi
\rho_\sigma(\epsilon) |V(\epsilon)|^2$. Since the coupling to a
continuum of states in a quantum impurity model is fully defined by its
hybridization function $\Gamma(\epsilon)$, we replace
$\sqrt{\rho_\beta(\epsilon)} V(\epsilon)$ by
$\sqrt{\Gamma_\beta(\epsilon)/\pi}$ in the following.

We now introduce the combinations of states
\begin{equation}
\label{sep}
\begin{split}
g_{\epsilon\sigma} &= \frac{1}{\tau} \sum_m \gamma_m^+ c_{\epsilon m \sigma}, \\
h_{\epsilon\sigma} &= \frac{1}{\tau} \sum_m \gamma_m^- c_{\epsilon m \sigma}, \\
\end{split}
\end{equation}
where the normalization factor $\tau$ is defined as
\begin{equation}
\tau = \left[ \sum_m \left| \gamma_m^\pm \right|^2 \right]^{1/2} =
1/\sqrt{2}.
\end{equation}
The two sets $g$ and $h$ are (canonical) fermionic operators,
$\{g_{\epsilon\sigma},g^\dag_{\epsilon'\sigma'}\}=\delta(\epsilon
-\epsilon')\delta_{\sigma\sigma'}$ and likewise for $h$, and they are
orthogonal to each other: $\{ g_{\epsilon\sigma},
h_{\epsilon'\sigma'}^\dag \}=0$.

We obtain
\begin{equation}
\begin{split}
H_0 &= 
\int_0^D d\epsilon\ \epsilon\ g^\dag_{\epsilon\uparrow}
g_{\epsilon\uparrow} + 
\int_{-D}^0 d\epsilon\ \epsilon\ g^\dag_{\epsilon\downarrow}
g_{\epsilon\downarrow} \\
&+ \int_0^D d\epsilon\ \epsilon\ h^\dag_{\epsilon\uparrow}
h_{\epsilon\uparrow} + 
\int_{-D}^0 d\epsilon\ \epsilon\ h^\dag_{\epsilon\downarrow}
h_{\epsilon\downarrow},\\
H_\mathrm{hyb} &= \int d\epsilon\ \sqrt{\Gamma(\epsilon)/\pi}
\ \tau \\
&\times \Bigl( 
\theta_\uparrow(\epsilon) d^\dag_{\uparrow} h_{\epsilon\uparrow}
+
\theta_\downarrow(\epsilon) d^\dag_{\uparrow}
h_{\epsilon\downarrow} \\
&+
\theta_\uparrow(\epsilon) d^\dag_{\downarrow} g_{\epsilon\uparrow}
-
\theta_\downarrow(\epsilon) d^\dag_{\downarrow} g_{\epsilon\downarrow}
\Bigr) + \mathrm{H.c.}
\end{split}
\end{equation}
We introduced
$\Gamma(\epsilon)=\Gamma_\uparrow(\epsilon)+\Gamma_\downarrow(\epsilon)$,
which is non-zero for all $\epsilon$ except at $\epsilon=0$, where
there is a linear pseudo-gap in the density of states. In $H_0$ we
omitted writing those combinations of the conduction-band states which
do not couple to the impurity, as they play no role in the following.

We note that the impurity spin-up orbital only couples to the $h$
states, and the impurity spin-down orbital only to the $g$ states.
Since the $g$ and $h$ states are only defined for the combinations
$\epsilon\geq0$, $\sigma=\uparrow$ and $\epsilon\leq 0$,
$\sigma=\downarrow$, we may actually drop the spin label and write
\begin{equation}
\begin{split}
g_{\epsilon} &= \theta_\uparrow(\epsilon) g_{\epsilon\uparrow}
- \theta_\downarrow(\epsilon) g_{\epsilon\downarrow},\\
h_{\epsilon} &= \theta_\uparrow(\epsilon) h_{\epsilon\uparrow}
+ \theta_\downarrow(\epsilon) h_{\epsilon\downarrow}.
\end{split}
\end{equation}
Thus we are dealing with two channels of spin-less electrons (defined
for all $\epsilon$) with a pseudo-gap density of states.
The Hamiltonian is then
\begin{equation}
\begin{split}
H_0 &= \int_{-D}^D d\epsilon\ \epsilon\ \left( 
g^\dag_{\epsilon}g_\epsilon
+ 
h^\dag_{\epsilon}h_{\epsilon}
\right), \\
H_\mathrm{hyb} &= \int_{-D}^D d\epsilon\ \sqrt{\Gamma(\epsilon)/\pi}
\ \tau 
\ \left(
d^\dag_{\uparrow} h_\epsilon + 
d^\dag_{\downarrow} g_\epsilon
\right)+\mathrm{H.c.}
\end{split}
\end{equation}
As a final step, the particle type label $h$ and $g$ can be replaced
by a (pseudo)spin label, i.e., $b_{\epsilon\uparrow} =
h_{\epsilon}$ and $b_{\epsilon\downarrow} = g_{\epsilon}$. 
The operators $b_{\epsilon\sigma}$ are canonical fermionic operators.
The final Hamiltonian takes the form of the regular single-impurity
Anderson model (compare with Ref.~\onlinecite{krishna1980a}):
\begin{equation}
\begin{split}
H_0 &= \sum_\sigma \int_{-D}^D d\epsilon\ \epsilon\ b^\dag_{\epsilon\sigma}
b_{\epsilon\sigma}, \\
H_\mathrm{hyb} &= \sum_\sigma \int_{-D}^D d\epsilon\
\sqrt{\Gamma(\epsilon)/\pi}\ \tau\ d^\dag_{\sigma} b_{\epsilon\sigma}
+ \mathrm{H.c.}
\end{split}
\end{equation}

The end-result of the derivation is thus trivial in its form, as the
low-energy effective model corresponds to the conventional Anderson
impurity model (with a pseudo-gap in the hybridization function).
Nevertheless, it should be noted that the operators
$b_{\epsilon\sigma}$ correspond to a rather complicated combination of
the original conduction-band electron states. Recently, formally the
same effective model has been studied in the context of Anderson
impurities adsorbed on a graphene layer \cite{cornaglia2009}. It is
interesting to note that despite the differences between the physical
content of the Hamiltonian in these two different cases (as already
observed in Refs.~\onlinecite{liu2009} and \onlinecite{feng2009}) the
effective impurity model in a suitably transformed basis is the same.
It should also be noted that this reduction is only possible starting
from an Anderson-type impurity model. Starting from a Kondo-type
impurity model, $H_K=J \vc{S}\cdot(\sum_{kk'\alpha\beta}
\psi^\dag_{k\alpha} \boldsymbol{\sigma}_{\alpha\beta}
\psi_{k'\beta})$, we would end up with an effective model with an
infinite number of channels (index $m$), because there is no
equivalent of Eq.~\eqref{sep} to perform a separation of the linear
combinations of states. If a spin-only model is desired, the
Schrieffer-Wolff transformation has to be performed as the last step
of the derivation.

For orbitals which couple with the continuum via $V_\vck$ which
depends on the azimuthal angle, a similar result would ensue. For
example, for $d_{xz}$ and $d_{yz}$ we have $V_\vck=V_k \cos\phi$ and
$V_\vck=V_k \sin\phi$, respectively, while for $d_{x^2-y^2}$ and
$d_{xy}$ we have $V_\vck=V_k \cos2\phi$ and $V_\vck=V_k \sin2\phi$,
respectively. It easy to see, however, that the angular additional
factors corresponds to a mere shift of $m$ by 1 or 2 in
Eq.~\eqref{cpm}, thus we end up with the same effective model,
although with different combinations of continuum states coupling to
the impurity. Of course, the Fermi surface/line is not axially
symmetric either, but rather reflects the symmetry of the lattice (for
example, due to warping effects resulting from the higher-order
spin-orbit coupling terms, the Fermi surface in the case of
Bi$_2$Te$_3$ has a hexagram/hexagon shape and becomes circular only in
the vicinity of the Dirac point \cite{fu2009, chen2009}). Again, this
only affects the combinations of states which couple with the impurity
orbital, however it does not change the results in a qualitative way.

The properties of the Anderson impurity in pseudo-gap Fermi baths are
well known \cite{bulla1997,gonzalez1998}: for a linear pseudo-gap,
$\rho \propto |\omega|$, the system flows to a local-moment (LM) fixed
point where the impurity is effectively decoupled from the conduction
states and remains unscreened at low temperatures, unless the
hybridization $\Gamma$ is strong enough. The LM fixed point for
$\epsilon_F=0$ is very resilient to various perturbation; in
particular, it persists for a large range of parameters away from the
particle-hole symmetric point \cite{gonzalez1998}. With $\epsilon_F$
away from the Dirac point, however, the density of states is finite
and the impurity is screened in the conventional Kondo effect, thus
the system flows to the strong-coupling fixed point. In
topological insulators, unlike in graphene, there is no a-priori
reason for the Fermi level to sit at the Dirac point,
thus the impurity will typically undergo Kondo screening at some
low-enough temperature $T_K$. The impurity contribution to the total
entropy $S_\mathrm{imp}$ is 0 much below $T_K$ \cite{wilson1975,
krishna1980a, bulla2008}, which implies that despite a complex
structure in the spin/reciprocal space, no anomalies are expected in
the low-temperature thermodynamics (including the magnetic
susceptibility): the impurity spin is fully screened (not
one-third screened \cite{feng2009}). This does not preclude a
complex structure of the Kondo cloud, which might indeed exhibit
non-trivial spatial and spin dependence. It is, however, unclear
whether these features could ever be directly observed, since the
Kondo cloud is rather elusive.

An Anderson-model-type magnetic impurity on the surface of a
topological insulator may be fully Kondo screened. For temperature
much below the Kondo temperature $T_K$, the backscattering of the
surface state electrons will be prohibited, as in the case of
non-magnetic adsorbates. No opening of the gap in the surface-state
band is expected in this regime. At temperatures above $T_K$, however,
the spin-flip scattering events will connect the points on the
opposite sides of the Fermi surfaces. The possible opening of the gap
might be, however, masked be thermal smearing effects. On the other
hand, for ferromagnetically Kondo coupled impurities or for large-spin
impurities with easy-axis magnetic anisotropy, Kondo screening is
non-effective and strong scattering effects may indeed be expected.
Further work should explore the behavior of multiorbital impurity
models and take into account the full spin and orbital structure of
the hybridization parameters $V_{k\alpha,m\beta}$, where $m$ is the
impurity orbital index, while $\alpha$ and $\beta$ are the spin
indexes of surface and impurity states, respectively.

\begin{acknowledgments}
R.Z. acknowledges the support of the Slovenian Research Agency
(ARRS) under Grant No. Z1-2058.
\end{acknowledgments}

\bibliography{topo}

\end{document}